\documentstyle[aps,prl,multicol,graphics,epsfig]{revtex}
\begin{document} 

\draft 
\flushbottom

\title{Resonant radiation pressure on  neutral particles in a waveguide}

\author{R. G\'{o}mez-Medina$^*$, P. San Jos\'{e}$^*$, 
A. Garc\'{\i}a-Mart\'{\i}n$^{*\ddagger}$,
M. Lester$^*$\cite{Marcelo},
M. Nieto-Vesperinas$^\dagger$ \&
J.J. S\'{a}enz$^{*\ddagger}$\cite{corresp}}

\address{$^*$ Departamento de F\'{\i}sica de la Materia Condensada, 
         Universidad Aut\'{o}noma de Madrid, E-28049 Madrid, Spain.\\
         $^{\dagger}$ Instituto de Ciencia de Materiales, CSIC,
         Campus de Cantoblanco, E-28049 Madrid, Spain. \\
         $^\ddagger$  Instituto de Ciencia de Materiales ``Nicol\'{a}s
Cabrera'', Universidad Aut\'{o}noma de Madrid, E-28049 Madrid, Spain.}

\date{\today} 
\maketitle 

\begin{abstract}  

A theoretical analysis of electromagnetic forces on neutral particles in
an hollow waveguide is presented. 
We show that the effective scattering cross section of a very small
(Rayleigh) particle can be
strongly modified inside a waveguide.  
The coupling  of the scattered dipolar field with the waveguide modes induce a
resonant enhanced backscattering state of the scatterer-guide system close
to the onset of new modes.  The particle effective cross
section can then be as large as the wavelength even far from any transition
resonance.
As we will show, a small particle  can be strongly
accelerated along the guide axis while being
highly confined in a narrow zone of the cross section of the guide.

\end{abstract} 

\pacs{42.50.Vk, 32.80.Lg, 42.25.Bs}

\begin{multicols}{2} 
Demonstration of levitation and trapping of micron-sized particles  by
radiation pressure dates back to 1970 and the experiments reported by
Ashkin and co-workers \cite{ashkin1}. Since then, manipulation and 
trapping  of neutral particles by optical forces has had a revolutionary
impact on a variety of fundamental and applied studies in physics,
chemistry and biology \cite{ashkin23}. These ideas were extended to
atoms and molecules where radiation pressure can be very large due to the
large effective cross section (of the order of the optical wavelength) at
specific resonances\cite{ashkin1,Letok}.  When light is tuned close to a
particular transition, optical forces involves (quantum) absorption  and
reradiation by spontaneous emission as well as coherent  (classical)
scattering  of the incoming field with the induced dipole \cite{Askar}.
Selective control of the strong interplay between these two phenomena is
the basis of laser cooling and trapping of neutral atoms 
\cite{Chu}. 

However, far from resonance, light forces on
atoms, molecules and nanometer sized particles are,  in general, very
small.   Here we show that the scattering cross section of a very small
(Rayleigh) particle can be strongly modified inside a waveguide. 
The coupling  of the scattered dipolar field with the waveguide modes induce a
resonant enhanced backscattering state of the scatterer-guide system close
to the onset of new modes.  Just at the resonance, the effective cross
section becomes of the order of the wavelength leading to an enhanced
resonant radiation pressure which does not involve any photon absorption
phenomena. As we will show, a small particle  not only can be strongly
accelerated along the guide axis  but it can also
be highly confined in a narrow zone of the cross section of the guide.

For the sake of simplicity we consider a two-dimensional $xz$ waveguide
with perfectly conducting walls and cross section $D$. The particle is then
represented by a cylinder located at $\vec{r_0}=(x_0,z_0)$
with its axis along $oy$ and radius much smaller
than the wavelength (see top of Fig.1). However, apart from some 
depolarization effects, the
analysis contain the same phenomena as the full three-dimensional problem
\cite{Lester} and hence it permits an understanding of the basic
physical processes involved in the optical forces without loss of
generality.

An s-polarized electromagnetic wave is assumed (the electric field parallel to
the cylinder axis), 
$\vec{E}(\vec{r}) = \exp(-i \omega t) E^0(\vec{r}) \vec{y}$
with wavevector $k=\omega/c=2\pi/\lambda$. 
For a single-mode waveguide ($D/2 < \lambda \le D$),  the
incoming electric field can be written as the sum of two interfering plane
waves: $
E^0(\vec{r}) =   E_0  \left(\exp(ik_z z + ik_x x) - \exp(ik_z z - ik_x x)\right)
$
where
$k_{z}=k \cos(\theta)$, $k_{x}=k \sin(\theta)= \pi/D$.
The scatterer 
can be characterized by the scattering phase shift $\delta_0$ or by its
polarizability $\alpha$
and Rayleigh scattering cross section $\sigma$ \cite{libro,phase}.
The time average force $\vec F$  can be written as the sum of an optical
gradient force and a scattering force:
\cite{Gordon}:
\begin{equation} \label{fuedelta}
 \vec F  = \left\{ \frac{1}{4} 
\alpha \vec \nabla |E^{inc}|^2 + \frac{ \langle \vec S \rangle}{c} \sigma \right
\}_{\vec{r}=\vec{r}_0} 
\end{equation}
where $E^{inc}$ is the total incident field on the particle and 
$\langle \vec S \rangle$ is the time average Poynting vector. 

If we neglect the multiple scattering effects between the scatterer and the
waveguide walls, the interaction would be equivalent to that of a particle
placed in the interference pattern of two crossed plane wave beams.  
In this case $E^{inc} = E^0$ and the
theory of radiation pressure in a waveguide is straightforward. The
longitudinal force $F_z^0$ (per unit length) 
can be written in terms of the 
average power density of the incident beams, 
$ \langle S \rangle = \epsilon_0 c |E_0|^2$,  as
\begin{equation}
F_z^0 = 
 2  \sigma \epsilon_0 |E_0|^2  \cos(\theta) 
\sin^2(\frac{\pi x_0}{D})
\end{equation}
which is maximum ($F^0_{z_{max}} = 2  \sigma \epsilon_0 |E_0|^2  \cos(\theta)$) 
just in middle of the waveguide.
The transversal force 
induces an optical potential along
$x$ given by:
\begin{equation}
U^0_x= -  \alpha |E_0|^2 \sin^2(\pi x_0/D).
\end{equation}
which, for $\alpha > 0$, confines the particle near the center of the
waveguide.
This is the two-dimensional analogue of 
previous approaches on laser-guiding of atoms and particles in hollow-core 
optical fibers
\cite{OlShanii,Renn1,Renn2} where 
the interaction of the dipole field with the guide walls was neglected.

The scattering with the waveguide walls may induce however a dramatic effect 
on the optical forces on the particle.
The scatterer radiates first a dipole field generated by the  field of the
incoming mode
$E^0$. Then, the scattered field, perfectly reflected by the waveguide
walls,
goes back to the scatterer  changing the field inciding on the
scatterer and so on.  This multiple
scattering process can be regarded as produced by a set of infinite image
dipoles \cite{chu1,Kunze}.
From the exact solution for the total field together with
equation (1), we found that, for a single mode waveguide, the forward component 
of
the force $F_z$ can be written in terms of the waveguide transmittance $T$ 
(defined as the ratio between the outgoing and incoming energy flux; 
$0 \le T \le 1$) as:
\begin{equation}
F_z = 2  D \epsilon_0 |E_0|^2  \cos^2(\theta)  (1-T(x_0))
\end{equation}
where $T$ depends on the transversal position of the particle $x_0$. 

The transmission coefficient $T$ had been discussed before in the context
of electronic conductance of quasi-one-dimensional
conductors with point-like attractive 
impurities \cite{chu1,Kunze,Bagwell,Tekman}. $T$ presents two peculliar
properties: {\em i)} when $D/\lambda$ is just
at the onset of a new propagating mode, the scatterer becomes transparent;
{\em ii)} interestingly, when $D/\lambda$ is close
but still below a mode threshold, the transmittance of a single mode 
waveguide presents a dip down to exactly $T=0$. 
This  backscattering resonance, that was associated to the existence of a
quasi-bound state induced by the attractive impurity
\cite{Bagwell,Kunze}, can be achieved for  any attractive scattering
potential of arbitrarily small strength \cite{Kunze}, i.e. {\em for
arbitrarily small 
polarizability $\alpha$ and cross section $\sigma$} \cite{phase}.
This resonance has a
pronounced effect on the radiation forces.

In Fig. 2 we plot the transmission coefficient as a function of both the
waveguide width, $D/\lambda$, and the scatterer position $x_0$ for 
$\delta_0=10^o$. 
Near the threshold of the second mode ($D/\lambda \lesssim 1$),
 $T$ presents sharp dips (down to $T=0$) at some particular 
positions of the scatterer.
Just at the enhanced backscattering resonances, the longitudinal
force present  strong maxima $F_{z_{max}}$ 
which can be compared with $F^0_{z_{max}}$ (the maximum
force obtained neglecting the interactions with the walls)  
$F_{z_{max}}/F^0_{z_{max}}= \cos(\theta) D/\sigma$ (see Fig. 3), 
i.e. at resonance the
interaction cross section of the particle in the waveguide
can be as large as the total cross
section of the waveguide independently of its value $\sigma$ in
the unbounded space. The force enhancement factor can be huge: 
for $\delta_0 = 10^o$ ($\sigma \approx 20 nm$ ) and at micron wavelengths, 
it is of the order of 50,  while for a
nanometer scale particle ($\delta_0 \approx 2^o$) 
this enhancement would be $\approx 10^3$ 
for two-dimensions, but  $\approx 10^6$ in true three-dimensional systems!
This result is rigurously true for perfect walls.
In actual waveguides however, the radiation losses through the walls and
the scattering with surface defects \cite{break1}
may modify the resonace behaviour for $\sigma$ values  comparable to the
surface roughness.

The resonance is related to the strong coupling between the incoming mode
and the first evanescent mode in the waveguide. This can be seen by
plotting the field intensity inside the waveguide for different particle
positions (Fig. 1). At the resonance, the field around the particle
corresponds to that of the second mode in the waveguide (with a node
in the waveguide axis) which decay far from the defect. 
When the particle is located at the middle of the waveguide,
there is no coupling of the scattered field with the first evanescent mode
and $F_z$ presents a minimum. 

Transversal forces are also strongly affected by the resonances. 
Although the main contribution to these forces come from polarization
effects (i.e. proportional to $ \nabla |E^{inc}|^2$),
in contrast with the free space case, the lateral forces  have also a
contribution of pure  scattering  origin due to the reflections of the
flux from the walls. 
The induced transversal confining potential far from the mode threshold
presents a single well similar to that discussed for the unbound system.
However, near the 
resonance condition it  presents two strong minima reflecting the excitation of
the evanescent mode. In Fig. 3 we plot the normalized 
longitudinal force $F_z$ and  transverse confining potential $U_x$.
The particle will be strongly confined in a small region inside the
waveguide where the forward longitudinal force is maximum.
For example, for
$\delta_0 = 10^o$  and at micron wavelengths, the potential well is
more than one order of magnitude deeper than that obtained for the unbound
system.

In summary, we have discussed the electromagnetic forces on small neutral
particles in a hollow waveguide. In contrast with standard resonance
radiation forces, the waveguide-particle backscattering resonances 
discussed here 
do not involve  photon absorption processes and, we believe, open
intriguing posibilities of atom and molecule manipulation. 
Specifically, the depth of the potential wells for the particle in resonant
conditions and its remarkably large cross section suggest stable guiding
of the particle along the waveguide with extremely large accelerations.

We thank R. Arias, P. C. Chaumet, L. Froufe, F. J. Garc\'{\i}a-Vidal,
R. Kaiser, T. L\'{o}pez-Ciudad and L. Mart\'{\i}n-Moreno for discussions.
Work of M.L. was supported by a postdoctoral grant of the 
Comunidad Aut\'{o}noma de Madrid.
This work has been supported by the Comunidad Aut\'{o}noma de Madrid and
the DGICyT through Grants  07T/0024/1998 and  
No. PB98-0464.

\pagebreak

\begin{figure}
\narrowtext
{\begin{center}%
\psfig{file=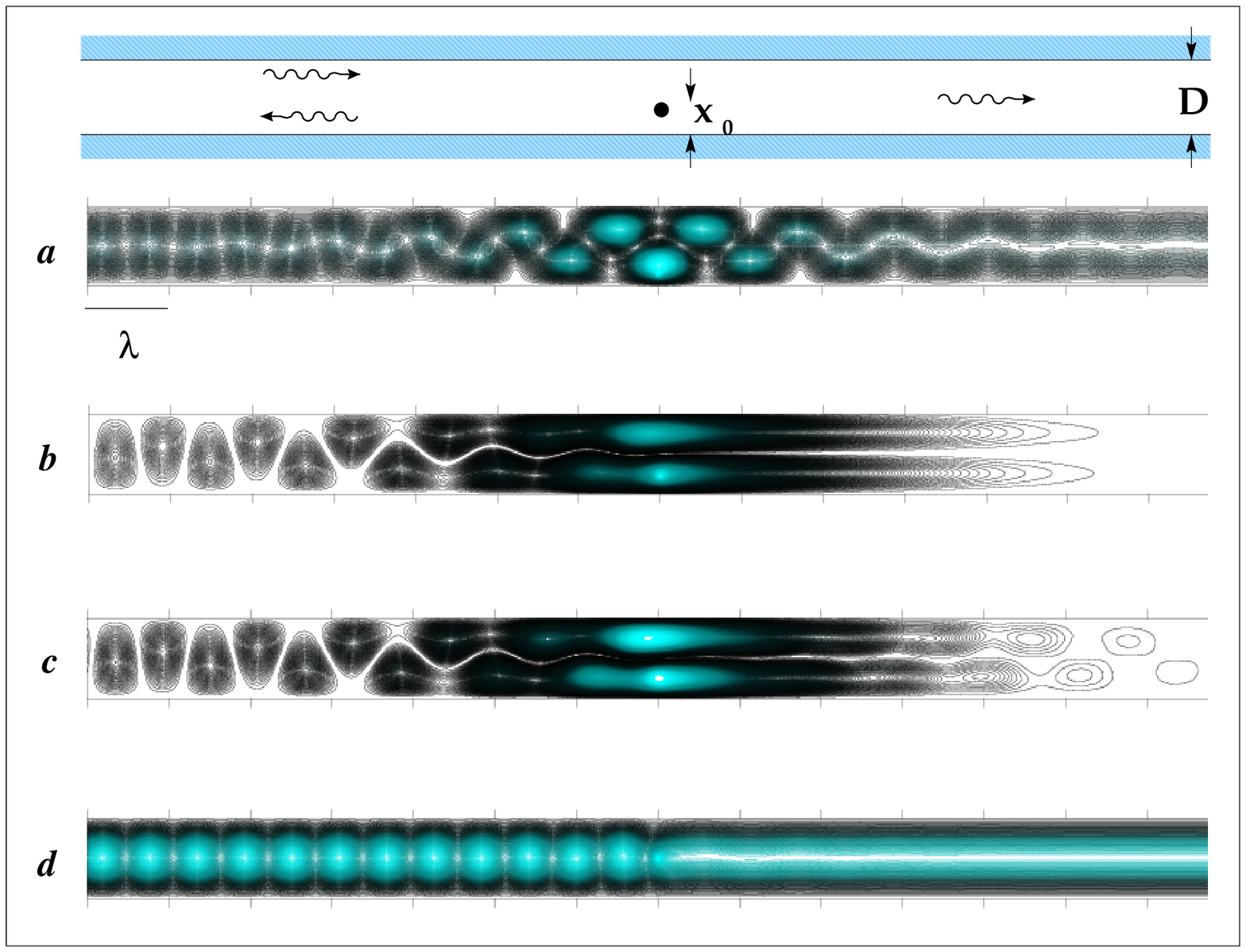,%
width=0.5\textwidth,clip=}
\end{center}}
\caption{Top: Sketch of the particle-waveguide system. 
Field intensity plots for different particle positions $x_0$ 
across the waveguide
({\em a)} $x_0/D=0.15 $, {\em b)} $x_0/D=0.22$, 
{\em c)} $x_0/D=0.25 $, {\em d)} $x_0/D=0.5$). The field incides from the
left side. }
\label{Figure 1.}
\end{figure}

\begin{figure}
\narrowtext
{\begin{center}%
\psfig{file=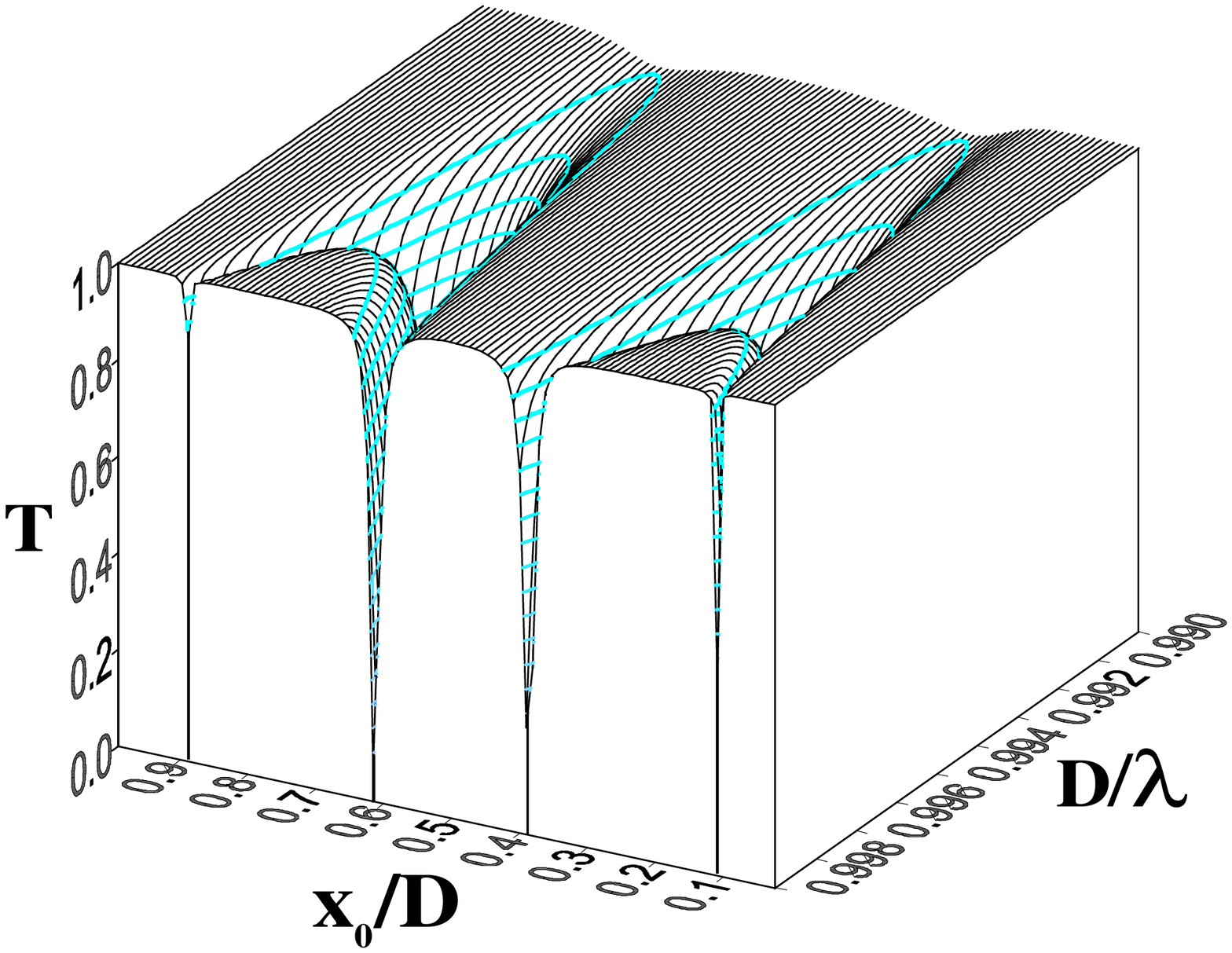,%
width=0.5\textwidth,clip=}
\end{center}}
\caption{Transmittance of a single mode waveguide versus width $D/\lambda$ 
and scatterer position $x_0/D$ for a fixed phase-shift value
$\delta_0=10^o$.}
\label{Figure 2.}
\end{figure}

\begin{figure}
\narrowtext
{\begin{center}%
\psfig{file=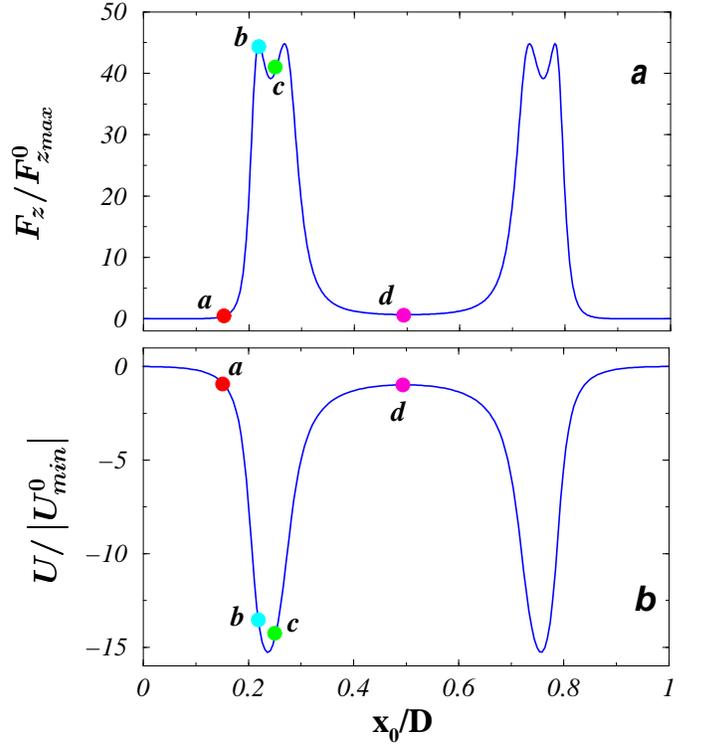,%
width=0.5\textwidth,clip=}
\end{center}}
\label{Figure 3.}
\caption{Longitudinal force $F_z$ {\bf (a)} and lateral confining potential $U$ 
{\bf (b)} versus particle position (normalized to the maximum force
$F^0_{z_{max}}$ and the minimum potential $U^0_{min}$ 
in the unbounded system). The values of $x_0/D$ corresponding to the dots
are those of 
({\em a, b, c, d}) in Fig. 1.}
\end{figure}

\end{multicols}

\begin{thebibliography}{99}
\bibitem{phase}
Assuming a s-polarized electromagnetic plane wave and
a cylindrical scatterer, the polarizability is related to the
scattering phase shift $\delta_0$ by
$\alpha = \epsilon_0 (2/k^2) \sin(2 \delta_0)$ .
A simple application of the optical theorem gives a
scattering cross section (cross length in this case) $\sigma =
(4/k) \sin^2(\delta_0)$.
For example,  for a dielectric rod of radius $a$,  dielectric 
constant $\epsilon$ and $ka \ll 1$,
$\delta_0 \approx \pi (\epsilon -1) (k a/2)^2$, 
$\alpha \approx  \pi \epsilon_0 (\epsilon -1) a^2 $ and
$\sigma \approx  \pi^2/(4k)  (\epsilon -1)^2 (ka)^4 $.
\end{thebibliography}
\end{document}